# Experimental and Numerical Analysis of Strain Gradient in Tensile Concrete Prisms Reinforced with Multiple Bars

Viktor Gribniak[1]*, Ronaldas Jakubovskis[1], Arvydas Rimkus[1,2], Pui-Lam Ng[3,4], David Hui[5]

[1]Laboratory of Innovative Building Structures, Vilnius Gediminas Technical University, Vilnius, Lithuania
[2]Laboratory of Composite Materials, Vilnius Gediminas Technical University, Vilnius, Lithuania
[3]Institute of Building and Bridge Structures, Vilnius Gediminas Technical University, Vilnius, Lithuania
[4]Department of Civil Engineering, The University of Hong Kong, Hong Kong, China
[5]Department of Mechanical Engineering, University of New Orleans, New Orleans, USA
*Corresponding author, Email: viktor.gribniak@vgtu.lt

**Abstract:** This work is a continuation of the ongoing research on deformation behavior of reinforced concrete elements under tension. The previous studies have revealed that deformation behaviors of elements reinforced with multiple bars and the traditional prismatic members reinforced with a center bar are essentially different. The latter layout, though typical of laboratory specimens, could not represent the norm of structures in real-life. Thus, a new test methodology to investigate the strain distribution in concrete prismatic members reinforced with multiple bars subjected to axial tension is devised. Prismatic concrete specimens with different reinforcement configurations were fabricated and tested using the proposed setup. Deformation behavior of the specimens is modeled with a tailor-designed bond modeling approach for rigorous finite element analysis. It is revealed that the average deformations of the concrete could be different from the prevailing approach of average deformations of the steel, and are dependent on the reinforcement configurations. Therefore, the efficiency of concrete in tension should be carefully taken into account for rational design of structural elements. The study endorses promising abilities of finite element technique as a versatile analysis tool whose full potential is to be revealed with the advent of computer hardware.

# 1. Introduction

Testing of composite elements under direct tension is of fundamental importance to reveal the tension load response and cracking behavior of reinforced concrete (RC). Although the direct tension test of a concrete prism embedded with a single reinforcing bar is the most widely adopted experimental arrangement for such purpose [1], the test configuration does not perfectly mimic the real structural behavior [2]; moreover, there is no standardized test setup established to-date. Notwithstanding the apparent simplicity of the setup, it might be difficult to interpret the test results: the experimental evidence often contradicts to the general assumption of similarity between average strains in the reinforcement and concrete. Moreover, the traditional tests typically provide measurements of average deformations along the embedded reinforcing bar and over the concrete surfaces, which is an over-simplification of the actual distribution of strains in the concrete. This limitation restricts the accurate assessment of the deformation and cracking behavior of concrete tension members [2].

Under the assumption that all tension at the cracked section is carried by the reinforcement, i.e. neglecting the softening behavior of the concrete after cracking and considering the idealized crack pattern (regularly distributed and fully formed transverse cracks), the predicted width of the cracks would be constant throughout the section depth. This is not in accordance with the reality, where the crack width and the tensile strain are not constant throughout the cracked section as confirmed by physical testing. Contradicting the experimental evidence (Figures 1a and 1b), such over-simplified assumption does not enable the representation of actual distribution of the strains in the cover concrete over the cracked section, where the crack width would vary in the manner of a wedged shape [3,4]. To illustrate the variation of crack width over the concrete cover, the experimentally obtained crack widths reported by Borosnyói and Snóbli [4] are plotted in Figure 1a. The measured crack widths at the upper and lower concrete surfaces are denoted as $w_1$ and $w_2$, respectively. From the experimental results, $w_1 = 0.35$ mm and $w_2 = 0.45$ mm. Through the concrete cover, the measured crack width varied almost linearly from the concrete surface towards the reinforcing bar, as shown in terms of ratios of $w_1$ and $w_2$ in Figure 1a.

The current approaches in deformation analysis of RC members are commonly based on the assumption that only a part of concrete cross-section under tension can carry tensile loads [5]. This concrete part is referred to as "effective concrete area in tension". This area is schematically shown in Figure 1c. In the cracked RC element, concrete undergoes complex stress-strain states, the cross-section becomes non-planar due to formation of primary and internal conical (Figure 1b), also known as "Goto", cracks [6] and the corresponding bond stress transfer mechanism between concrete and reinforcement. In prevailing design approaches for the cracking analysis, the concrete is divided into two regions in resisting tension, namely the "effective" and "ineffective" regions [5]. The "effective" region is demarcated by the relative magnitude of tensile stress in the concrete. Typically, due to the transfer of stress between concrete and reinforcement through the bond action, the boundary of "effective" region manifests a parabolic shape, as illustrated in Figure 1c. For the sake of simplification in structural design, an idealized stress-strain behavior may be assumed, such that the two regions are delineated with respect to their volume proportionately. However, a number of studies [4,7-9] have revealed noticeable limitations of the "effective area" concept related with its inability of representing the effects of concrete cover, loading conditions, stress-strain state, and configuration of the unreinforced area. The main uncertainty in connection with this concept is related to the complicacy of measurement of actual strain distributions in the volume of cracked concrete. Consequently, the real stress distribution in concrete is not perfectly understood and simplifications have to be applied in cracking and deformation analysis of RC structures. This introduces errors to the structural analysis and design processes. However, a scientific methodology to rectify the deficiency has been in lack.

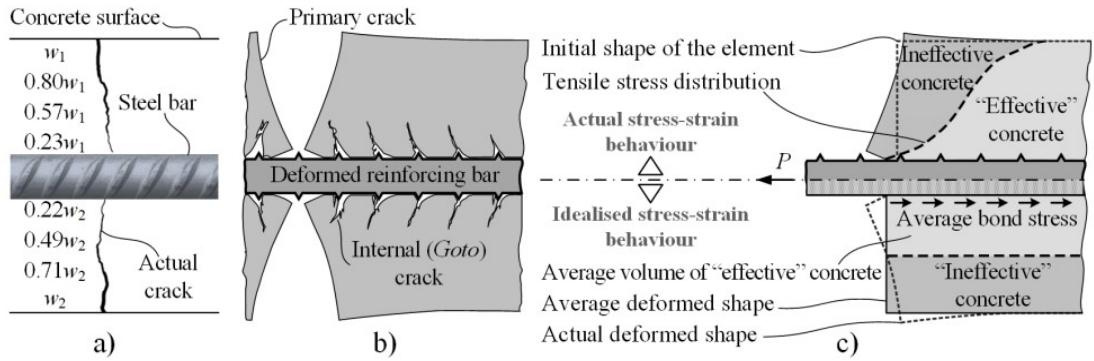

**Figure 1**. The concept of effective concrete in tension: (a) experimentally attained variation of the crack width [4]; (b) development of the internal cracks [6]; (c) effective concrete model

A number of techniques have been developed for the strain monitoring of RC specimens [10,11]. The most straightforward and commonly used method is measuring the displacement between two points to obtain average strain in the gauge length. Such specimens are commonly instrumented with linear variable displacement transducers (LVDT), which are attached to the concrete surface. Such equipment enables assessing average surface strains of the concrete. However, as mentioned at the beginning of this section, the deformations of the concrete surface and internal bar reinforcement might be different. For realistic analysis of the experimental behavior, the deformations at both locations must be monitored during the tests. Instrumentation arrangement for such purpose as well as cross-verification by computational analysis are among the objectives of this study, through which the strain gradient variations for tensile concrete prisms with different reinforcing bar arrangements are realistically reflected.

The average strains of the reinforcement might be identified by various means. For exposed sections of the reinforcing bars, attaching LVDT devices to the surface of bar is viable. Furthermore, a specimen also can be equipped with advanced monitoring systems such as internal gauging system [12,13] or optical sensors [14,15], which are suitable for precise assessment of the bar strains. In addition, the digital image correlation (DIC) technique is becoming an increasingly useful tool for tracking deformations at the concrete surface [16]. Modern image back-scattering techniques, such as X-ray [17], acoustic emission tomography [18], and magnetic resonance imaging [19], are available as indirect means of deformation measurement. However, the interpretation of data obtained from these non-contact methods is often complicated and may require users' judgement, and they are limited to simple specimen geometry and loading cases. Overall, the physical strain distribution within the RC member can be assessed only in an approximate manner.

A well-tailored numerical model could help solving this problem. Through an appropriate numerical approach, the deformation analysis is performed to reveal detail information about the strain distribution at the concrete surface and along the bar reinforcement. The numerical approach provides the ability to evaluate the intricate load transfer and internal cracking phenomena, and assess the effects on the structural behavior at member scale, which are too complex to be fully evaluated experimentally. It is able to investigate numerically the evolution of strain gradient in the concrete of a tensile element with load applied to the reinforcement. The study performed by Michou et al. [16] should be mentioned as a successful example of such analysis. However, this simulation approach has some limitations, as explained later in this paper.

To represent adequately the actual behavior of the element, the numerical model must be validated with respect to along reliable experimental data of representative specimens [2]. Broms and Lutz [7], Rostásy et al. [20], Hwang [21], Williams [22], and Purainer [23] have experimentally investigated deformation behavior (with particular attempt addressed to development of the cracks) of tensile elements with multiple bar reinforcement. It should be noted that the typical tensile elements with a bar at the central position might not be representative for all cases due to the unrealistically simplified loading condition: the load is applied directly to the bar, rather than the uniformly to the whole cross-section. Moreover, such elements are incapable to reflect the deformation behavior associated with group effects of closely spaced reinforcement bars [8,24,25]. Hence, the modeling results would not be representative of the behavioral characteristics in the tension zone of real structural members, where multiple bars exist.

This work continues series of publications by the authors [2,26,27] related to the analysis of deformation behavior of the tension elements. It was revealed that deformation properties of elements reinforced with multiple bars and the traditional prismatic members reinforced with a center bar are essentially different. The difference is closely related with the existence of a non-linear strain gradient in the concrete due to partial transferring of tensile stresses in single or group of bar reinforcement through the bond interaction mechanisms. Yannopoulos [28], Dominguez et al. [29], Tammo and Thelandersson [30], Michou et al. [16], and Gribniak et al. [2] have investigated this effect in concrete prisms reinforced with a center bar. The test program on concrete prisms reinforced with multiple bars carried out by the authors [31] revealed that the arrangement of reinforcement might alter the deformation behavior of nominally identical specimens. Counter-intuitive results were identified including the occurrence of negative (compressive) average strains at the concrete surface in some cases. The observed effect was related with a non-uniform distribution of strains in the concrete interacting with the bond under the externally applied load *P* (Figure 1c).

This paper investigates effect of arrangement of multiple bars on strain gradient in concrete prisms through an innovative methodological procedure that combines experimental and numerical (finite element) approaches. The experimental part of the methodology encompasses an innovative testing setup that secures the applied tension loading to be uniformly distributed among the multiple bars, as illustrated in Figure 2. The developed equipment allows to measure concrete surface strains at different locations and to monitor the deformation of each individual bar. Two series of the nominally identical prisms with different arrangement of the reinforcement were designed for this study. The prismatic specimens were reinforced either with four or with eight 10 mm bars distributed in 150 mm square section with different cover depths. The experimental data enabled calibration of the finite element models based on a regular bond model adapted from the literature. The model enables the evaluation of strain distributions in RC tensile elements.

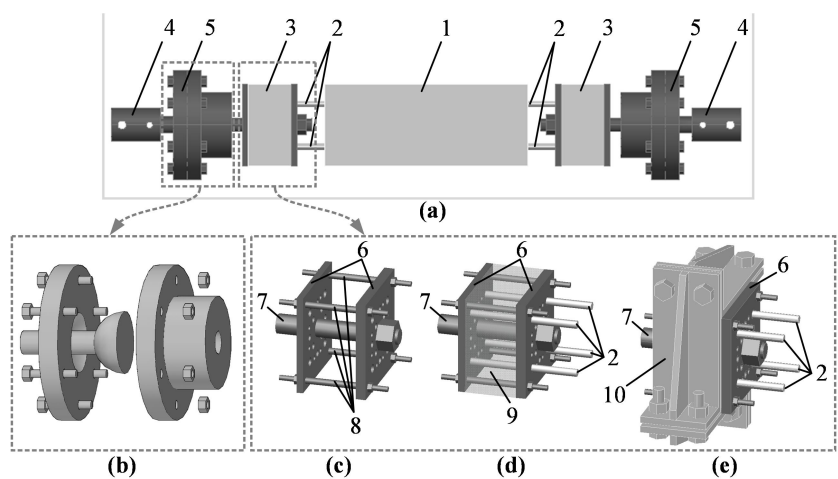

**Figure 2.** Equipment for testing concrete prisms reinforced with multiple bars: a) test specimen prepared for connecting the testing machine; b) spherical hinge; c) anchorage joint details; d) filled anchorage joint with installed bars; e) anchorage joint with steel brackets. The notations are following: 1 = concrete prism, 2 = reinforcement bars, 3 = anchorage joint, 4 = standard joint of a tension device, 5 = spherical hinge, 6 = steel plates of anchorage joint, 7 = central bar of anchorage

joint, 8 = supplemental bars of anchorage joint, 9 = concrete infill, 10 = supplemental equipment for shear restraint of the anchorage joint

**2. Description of the test setup**

The test procedure encompasses the basic principles described in the reference [31]. Design of the test equipment was governed by the idea of representing deformation behavior in the tensile zone of a realistic structural element. This arrangement aligns with the real case of multiple bars in the tensile concrete. A possibility of monitoring deformations of the reinforcement bars was another aim of the design. The test equipment is shown in Figure 2. As can be observed, the anchorage joints consist of two fixing plates articulated by a central bar that is connected to the tension device via a spherical hinge. The spherical hinge serves the purpose to reduce the possible eccentricity in applying the load due to imperfections arisen from inhomogeneity of the concrete and non-uniform formation of the cracks. The fixing plates are perforated (pre-formed with holes) for the reinforcement bars protruded from the concrete prism to pass through. The space between the two fixing plates is filled with concrete or other composite material. The filler material between the fixing plates allows each reinforcing bar to deform differently in the anchorage zone, thereby equalizing the stresses of the reinforcement bars and, accordingly, ensuring a central resultant force of the bar group. This choice of filler composition ensures the adaptive deformation of the anchor unit, as it allows control of the anchoring strength and displacements of the reinforcement bars. This reduces the physical eccentricity of the composite element and makes it possible to vary the material used for preparing the element. Steel brackets are used for restraining transverse deformations and providing additional confinement to the anchorage joints (Figures 2e). Either the concrete prism and anchorage blocks can be made of the same material and cast contemporarily or the anchorages with fixed reinforcement bars can be produced before casting the concrete element. The latter enables production of the pre-stressed elements. Average deformations and slip of the reinforcement can be monitored by utilizing the gap between the edges of the concrete prism and the anchorage blocks for instrumentation.

Numerous test trials were carried out for verifying the reliability of the developed equipment. A typical test setup is shown in Figure 3a. As can be observed, multiple linear variable displacement transducers (LVDT) were used for monitoring the average deformations of reinforcement bars, concrete surface, and total deformations of the equipment. The average deformations of the reinforcement were estimated using the LVDT devices attached to the bars very close to the surface of the concrete prism as shown in Figure 3b. Crack widths of the prism specimen were measured using a portable optical microscope (Figure 3c). LVDT devices were also used for identifying the slip of reinforcement as shown in Figure 3d. It was ascertained that the anchorage strength was sufficient to prevent bond failure of the reinforcement. The previous studies [2,31] have revealed the existence of strain gradient in the boundary zones of the concrete prisms with tensile load applied to the bar reinforcement. To evaluate this effect in the prisms reinforced with multiple bars, LVDT devices were positioned at different distances from the edge as shown in Figure 3e.

The tests were conducted with the employment of a 600 kN capacity servo-hydraulic machine under displacement control with a loading rate of 0.2 mm/min. The test setup is shown in Figure 4. In order to observe the development of cracks and cross-verify the LVDT readings, the frontal surface of the tension members was captured and processed by a digital image correlation (DIC) system. As shown in Figure 4, the images were captured using two high sensitivity digital cameras (Imager E-lite 5M). The cameras, incorporating a charge-coupled device (CCD) detector, have a resolution of 2456×2085 pixel at 12.2 fps (frame per second) frame rate. The cameras were placed vertically on a tripod at 2.5 m distance from the test specimens (Figure 4). The use of two cameras enabled reliable capturing of image data within their respective focal zone to minimize errors due to aberration. Multiple specimens with identical geometrical and mechanical properties were tested in prior for identifying the measurement errors. From the test results, the image data including the deformation extent, number of cracks and crack widths obtained from replicated specimens were generally consistent. The tests have proved the ability of the developed equipment to perform structural tests on tensile RC elements with precise measurement of strains in multiple points.

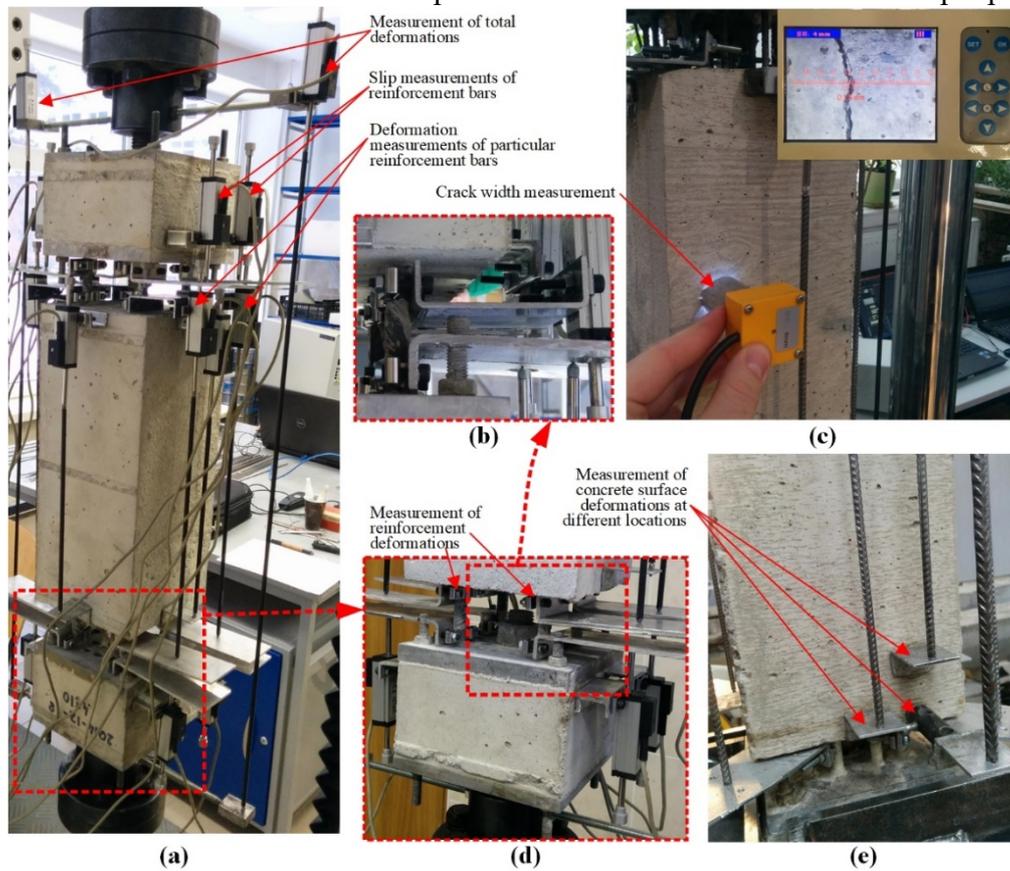

**Figure 3**. An innovative test layout: (a) the instrumented specimen; (b) LVDT devices attached to the bar for measurement of the slip (in respect to the anchorage block) and average deformation of the reinforcement; (c) crack measurement with optical microscope; (d) the slip measurement; (e) monitoring longitudinal deformations of concrete surface at different location of LVDT

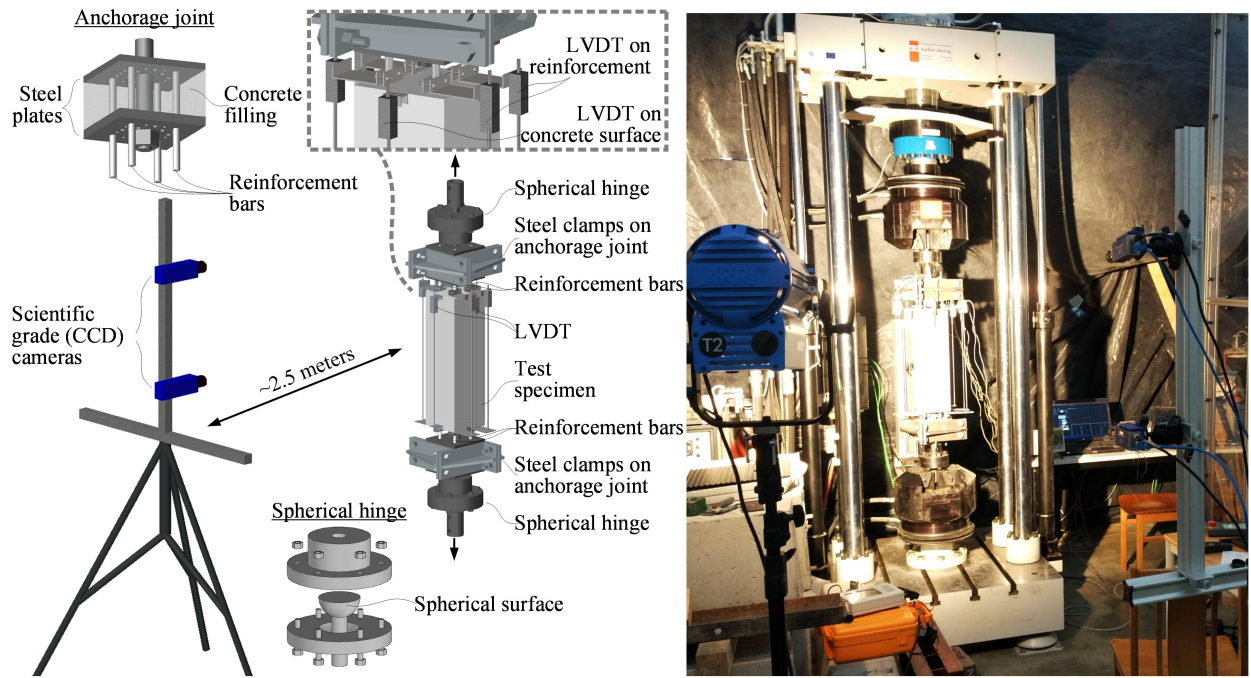

**Figure 4**. Monitoring deformations of concrete surface with LVDT devices and DIC system

## 3. Strain distribution in the concrete

Deformation behavior of RC during the crack formation stage is highly nonlinear. Goto [6] has related this process to the formation of secondary cracks due to transfer of bond stresses to the surrounding concrete between the transverse primary cracks. The internal secondary cracks redistribute strain fields within the concrete. To demonstrate this effect, ten RC tension members with identical 150×150 mm cross-section reinforced with four or eight 10 mm bars are included in the testing program. The corresponding reinforcement ratios are equal to 1.4% and 2.8%. The length of the concrete part (500 mm) is the same in all specimens. The bars are distributed in the section in a different manner. The test equipment described in the previous section is used.

The experimental data enable calibration of the finite element models with high precision to establish accurate numerical simulation for the evaluation of strain gradient in the concrete. The deformation problem is solved within the 3D strain domain employing non-linear constitutive models of the concrete and the bond with reinforcement reported in the literature [16,32]. The estimation of the strain gradient is based on the fundamental assumption that the numerical model, accurately representing the experimentally determined average deformations of the concrete surface and the bar reinforcement, is capable to predict the actual strain distribution in the concrete.

3.1 Experimental program

The specimens were cast in five batches. The concrete was laboratory-mixed and had a target compressive strength class of C30/37. A maximum 8 mm aggregate size was used. On the one hand, the influence of aggregate size on the mechanical properties of normal-strength concrete is relatively mild [33]. On the other hand, to represent the case of constructional concrete that typically contains coarse aggregates, the experimental program was not focused on the application of concrete with purely fine aggregates. Any sign of segregation problem was not observed during the concrete mixing and casting.

The compressive strength of the concrete was determined from $\varnothing 150 \times 300$ mm cylinders. All specimens were stored in water curing tank to minimize the drying shrinkage effect. All specimens were reinforced with 10 mm steel bars with the yielding strength $f_y$ and the elasticity modulus $E_s$ equal to 510.1 MPa and 199.5 GPa, respectively. The main characteristics of the test specimens are listed in Table 1. The tensile specimens were reinforced either with four or with eight 10 mm steel bars. The first digit in the notation of specimens designates the number of the reinforcement bars; the alphabet "*s*" indicates that the bars were made of steel; the number "10" corresponds to the bar diameter. The next alphabet (if any) designates a special arrangement of the reinforcement bars: "*c*" describes the cover enlarged from 30 mm (reference) to 50 mm; "*X*" designates a diagonal arrangement of the reinforcement bars, while "*R*" corresponds to the rectangular distribution of the bars. The last digit designates replications of nominally identical specimens. Cross-sections of the specimens are shown in Figure 5. As can be observed in Table 1, the prismatic specimens could be split into two series corresponding to the reinforcement ratio. The specimens of the first series (the reinforcement ratio $p = 1.4\%$) were designed in the same manner as in the previous study [31]. These prisms were produced for the purpose to analyze the cover effect on deformation behavior of the concrete, since the previous tests have revealed an extraordinary deformation behavior of the concrete surface related with an occurrence of average compressive strains at the concrete surface preceding formation of the first major (primary) crack [31]. The prisms of the second series ($p = 2.8\%$) were tested for the purpose to analyze the effect of closely spaced bars on development of the strain gradient in the concrete. Figure 6 shows the final crack patterns of all tested prisms. Different results are characteristic of the prisms with different cover depth. The comparative analysis, however, is possible only for the elements with the same reinforcement ratio, since the cracking layouts are corresponding to the different ultimate average strains of the reinforcement, $\varepsilon_m$.

**Table 1.** Parameters of the test specimens

| Batch | Specimen | Cover $c$, mm | Reinforcement ratio $p$, % | Compressive strength of concrete $f_{cm}$, MPa | Testing age, days |
|---|---|---|---|---|---|
| 1 | *4s10-1* | 30 | 1.4 | 46.7 | 16 |
| 1 | *4s10-2* | 30 | 1.4 | 46.7 | 16 |
| 2 | *4s10-3* | 30 | 1.4 | 44.7 | 15 |
| 2 | *4s10-4* | 30 | 1.4 | 44.7 | 15 |
| 3 | *4s10c-1* | 50 | 1.4 | 45.3 | 14 |
| 3 | *4s10c-2* | 50 | 1.4 | 45.3 | 14 |
| 4 | *8s10X-1* | 30/50[*] | 2.8 | 38.0 | 35 |
| 4 | *8s10X-2* | 30/50[*] | 2.8 | 38.0 | 36 |
| 5 | *8s10R-1* | 30 | 2.8 | 43.4 | 35 |
| 5 | *8s10R-2* | 30 | 2.8 | 43.4 | 42 |

[*]The cover depths corresponding the outer and the inner reinforcement frames are indicated.

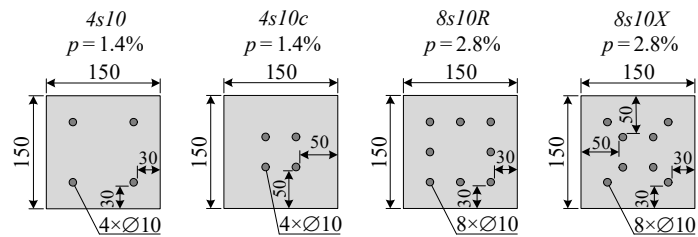

**Figure 5**. Cross-sections of the test specimens

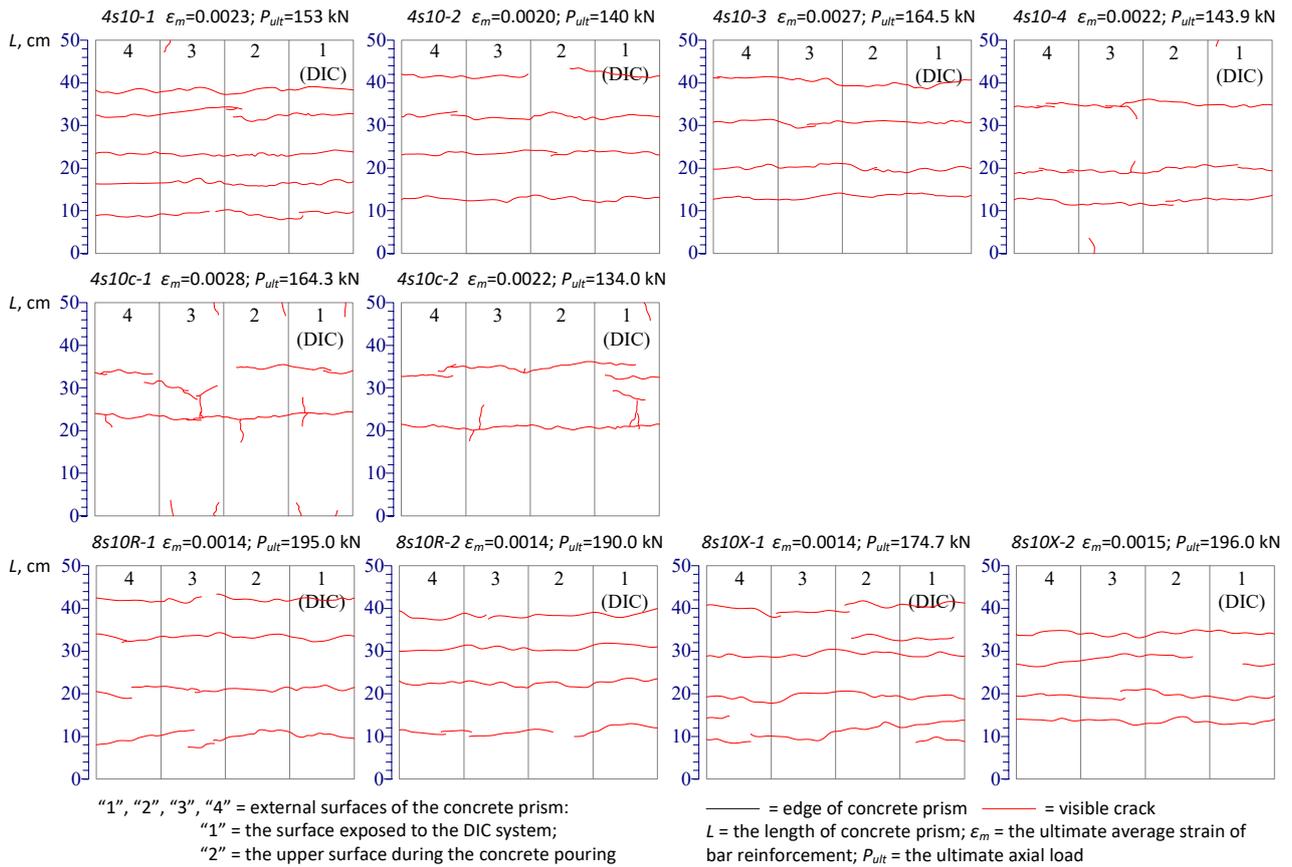

**Figure 6.** Final crack patterns. Note: these schemes represent crack projections to all four side-surfaces of the prisms

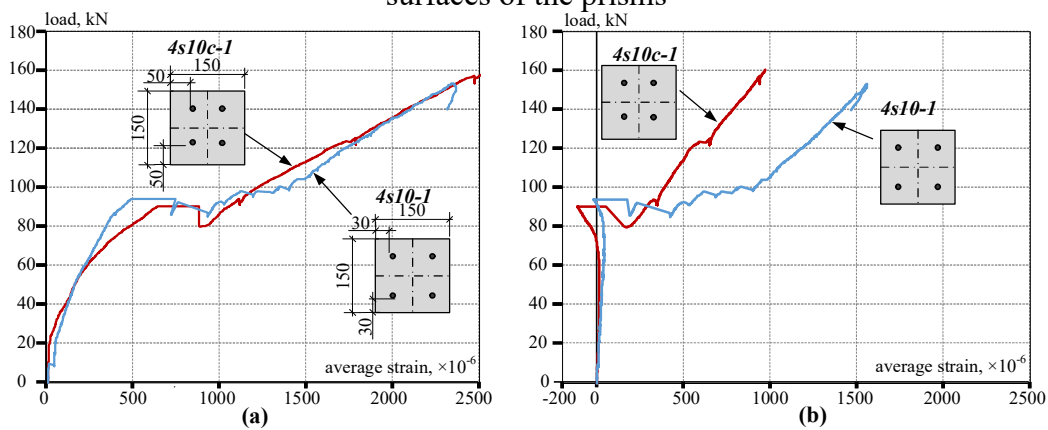

**Figure 7.** Experimentally estimated average strains of: (a) reinforcement and (b) concrete surface

The average strains the reinforcing bars and the concrete surface were assessed by using the monitoring results of the LVDT devices shown Figure 4. The average strain of the reinforcement was obtained by averaging the strain values of all bars, while the concrete surface deformations were averaged for all lateral surfaces of the concrete prism. The obtained diagrams of two selected elements with different cover thicknesses are shown in Figure 7. As can be observed, the average strains of the reinforcement and the concrete surface are apparently different. Figure 7a does not represent any noticeable differences between the diagrams of the specimens with different cover; however, the increased cover induces a "negative" deformation of the concrete surface at the loads approaching the cracking moment (Figure 7b). In this regard, it should be recalled that the concrete prisms were loaded by applying tension load to the reinforcement bars as shown in Figures 2-4. Such loading scheme causes deformation localization in the concrete prism transferring the bond stresses [16,30]. A relatively small reinforcement ratio (1.4%) of elements reinforced with four 10 mm bars in combination with decrease of the distance between the bars (of the prisms *4s10c*) might cause an increase of the concrete strains around the reinforcement bars in the boundary zones. That causes the end-surfaces of the prism to discord from planar form. The corresponding surface rotation about the boundary edge (Figure 1c) suppresses deformation of the side-surfaces. The contraction magnitude might exceed the tension deformations making average deformation of the concrete surface negative (compressive) as shown in Figure 7b. After the cracking, the differences between the strains of the concrete surface of the specimens with different cover increase drastically. Such a diverse behavior of the specimens with different cover shows that the respective strains in concrete surface and reinforcement should be dependent on the configuration of the reinforcement bars. Hence, comparative analysis of the results under the assumption of similarity of the strain of concrete surface and reinforcement may lead to misleading conclusions. Consequently, an experimental equipment that allows measuring strain only at the concrete surface does not enable assessment of actual distribution of strain within the concrete. This problem is tackled by means of a numerical approach for analyzing the strain distribution in the concrete, as discussed in the next section.

3.2 Numerical modeling

Numerical simulation has been carried out by using commercial finite element (FE) software ATENA [34]. The deformation problem was formulated based on the incremental constitutive law of materials. Following the symmetry conditions, the 1/16 part of the specimen was modeled in the three-dimensional domain as shown in Figure 8a that also describes the loading and boundary conditions. Isoparametric tetrahedral elements with 10 nodes and 4 integration points were used. The maximum element size assumed in the analysis was equal to 15 mm (the actual size was approximately equal to 13 mm) with five times refinement at the reinforcement and concrete interface. The finite element mesh example is shown in Figure 8b. Consequently, the numerical models of the prisms with 30 mm and 50 mm cover include respectively 40,000 and 35,300 finite elements. The numerical model was generated with provision of applying a simplified cylindrical bond model [16] with regular variation of the mechanical (strength and cohesion) characteristics representing effect of the ribs. Figure 8b schematically describes this modeling approach. Michou et al. [16] proposed mechanical parameters of the regular bond model of ribbed bars utilized in the present study. This simplification of the bar topology enables achieving a sufficient numerical accuracy to reproduce local behavior along a reinforcing bar, while significantly reducing the computational demands.

The bar is modeled as a cylindrical macroelement (with a constant diameter), while the contact interface with the concrete is modelled as a sequence of periodic variation of regular mechanical parameters. The assumed parameters of the bond model are described in Figure 8b, where $f_t$ and $C$ are the tensile and cohesion strength of the bond; $K_{nn}$ and $K_{tt}$ are the normal and tangential stiffness; $\varphi$ is the friction coefficient. Although the regular bond model is not suitable for representing the radial bond-stress component, Gribniak et al. [2] identified a satisfactory accuracy of such simplification for relatively small-diameter reinforcing bars (≤ 10 mm). This is because the magnitude of the radial component of bond stresses along small diameter bars is minor. For concrete, the SBETA model offered by ATENA is utilized. This model is based on the softening law of the cracked concrete proposed by Hordijk [32]. The necessary input parameters of the concrete, i.e. tensile strength, the modulus of elasticity, and the fracture energy, were assessed with *fib Model Code 2010* [35] using respective values of compressive strength from Table 1. The remaining parameters are assumed as default values described in ATENA. Reinforcement is modeled as a linearly elastic material. The experimental values of the compressive strength of the concrete and the elasticity modulus of the reinforcement are used in the model.

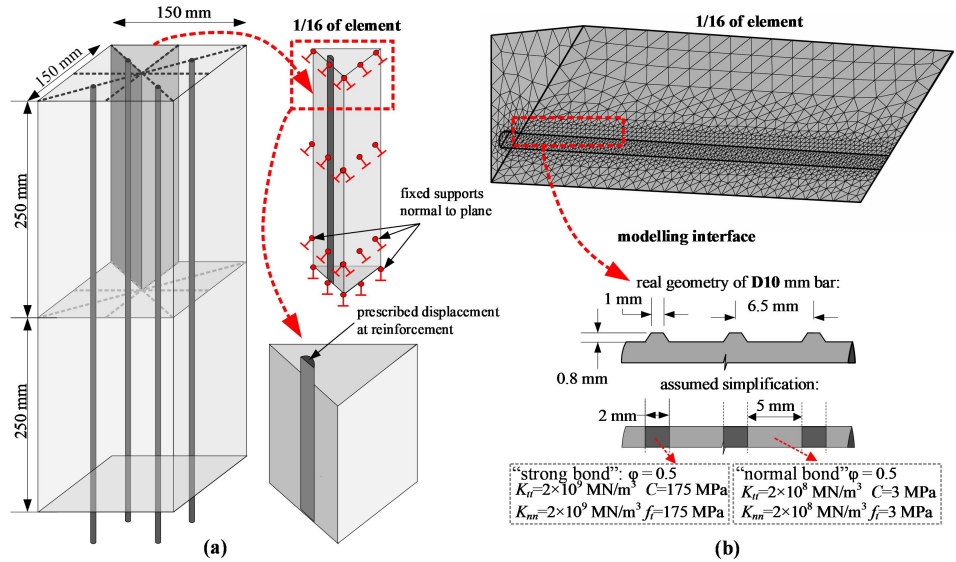

**Figure 8**. FE model: (a) the modeled segment and (b) FE mesh and bond modeling scheme

3.3 Discussion of the results

The load-deformation response of the prisms was simulated incrementally by applying deformations to the extremity surface of the reinforcement bar (Figure 8b). Figure 9 compares the simulated and experimentally identified load-average strain diagrams of the concrete surface (designated as "(C)") and the reinforcement (designated as "(R)"). This figure demonstrates a good agreement between the numerical predictions and experimental data (both deformations of the reinforcement and concrete surface) of the prisms reinforced with four bars (Figures 9a and 9b) and prisms reinforced with eight bars (Figures 9c and 9d).

A typical simulated strain distribution in the concrete of the prisms reinforced with four bars is shown in Figure 10. Strain value exceeding 0.003 was assumed as theoretical strain at crack (depicted as sky-blue regions in the strain map). Accounting for the assumed FE size (~13 mm), this stain value corresponds to the crack width visible at the concrete surface [36]. The regions with such strain levels are found to be localized in bands, which are in reasonably good agreement with the experimentally obtained crack patterns. The latter were identified at the same level of the applied deformations by using the DIC system.

An important result could be also related with "effectiveness" of the concrete sustaining the tensile stresses through the bond. The ATENA has predicted a different distribution of deformation in the concrete among the prisms with different cover. The strain distribution maps shown in Figure 10 can be referred for illustration. The predicted strains of the concrete below the theoretical cracking strain are shown in grey at these graphs. The volumes of the uncracked concrete are evidently different in the specimens with different cover. The diverse deformation behavior of the concrete is causing a different tension-stiffening effect characteristic of these specimens. The tension-stiffening effect can be estimated using the graphs shown in Figures 9a and 9b. It can be assessed as the difference between the theoretical "bare bar" response and the actual diagram of average strain of the reinforcement at a particular strain level. However, as shown in Figures 9c and 9d, the tension-stiffening effect differences are more significant in the prisms reinforced with eight bars. It is found that increase of the concrete cover from 30 mm to 50 mm would reduce the average deformations of the concrete approximately by 24%. Analysis of the numerically predicted strains of the concrete might help assessing the observed effects.

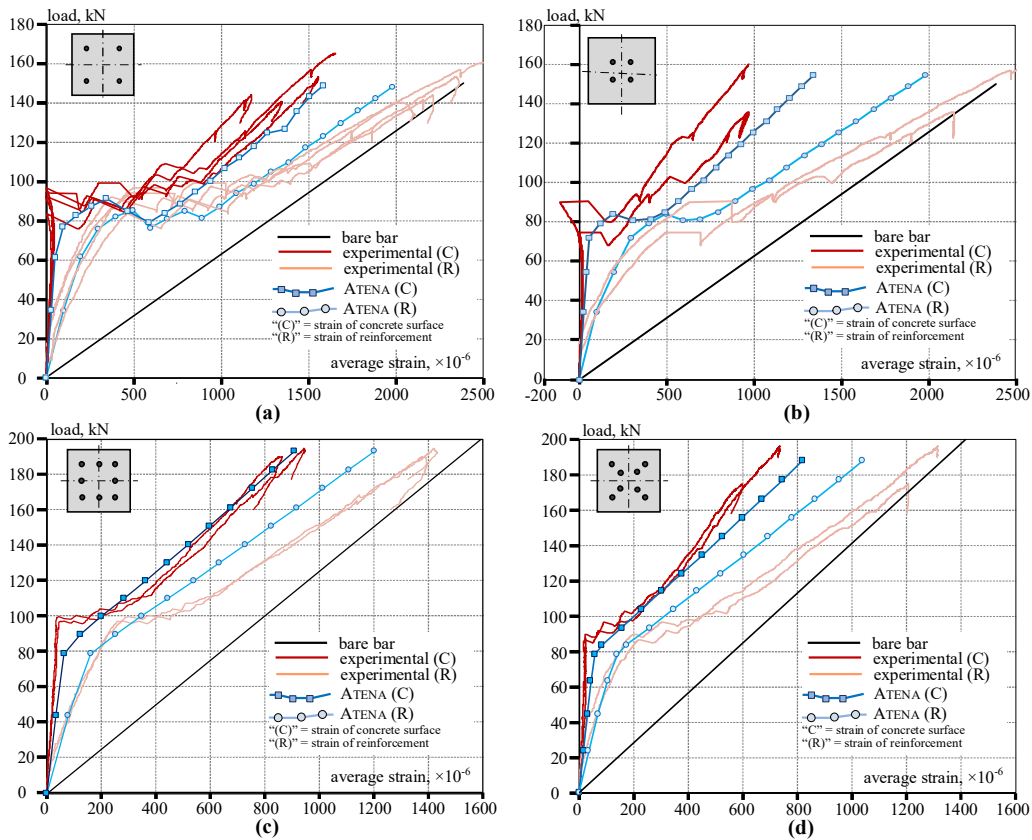

**Figure 9**. Load-strain diagrams determined at concrete surface and reinforcement bar of different prismatic samples: (a) specimen series *4s10*; (b) series *4s10c*; (c) series *8s10R* (c); (d) series *8s10X*

Figure 11 shows the strain distribution maps of concrete predicted by ATENA for different specimen configurations. Figure 11a presents distribution of tensile strains in the concrete in the same manner as it was done in Figure 10. The strain distributions in all specimens are related with the same average strain of the reinforcement, $\varepsilon_m$. Area of the concrete where the deformations are not exceeding the theoretical cracking strain is shown in grey, while the macro-cracks are indicated as sky-blue regions. It is evident that deformations of the concrete are closely related with the arrangement of the reinforcement bars. A good agreement of the predicted cracks with the experimental cracking patterns shown in Figure 6 could be identified as important outcome of these simulations. The strain distribution maps (Figure 11a) reveal that the bar arrangement effect decreases with increase of the number of the bars though it is remaining significant: the strain distribution of the prisms *8s10R* and *8s10X* is evidently different at all considered deformation levels.

An appearance of the negative (compressive) strains in concrete of the tensile elements, e.g., as shown in Figure 9b, can be explained by analyzing the strain distribution maps shown in Figure 11b. (This figure depicts the tensile strains in grey.) It is necessary to remind that the concrete is loaded by sustaining the tension deformations through the bond. However, as it can be observed in Figure 11b, the boundary region located at extremity of the concrete prism might accumulate a certain portion of the compressive strains due to an eccentric placement of the reinforcement bars relative to the external surface of the prism. In the prisms reinforced with four bars (*4s10* and *4s10c*), the eccentricity effect increase with the cover depth. A combination of the relatively low gradient of the tension strain (Figure 11a) and the compression deformations (Figure 11b) might cause an average negative deformation of the concrete surface. This effect decrease with the increase of the number of the reinforcement bars because of the increase of magnitude of the tension strains (Figure 11a). Notwithstanding an acceptable adequacy of the simulation results, a certain error remains characteristic of the deformation predictions (Figure 9). The error increases with number of reinforcement bars and/or with decrease of the distance between the bars. The prisms *8s10X* and *4s10c* could be considered as characteristic examples.

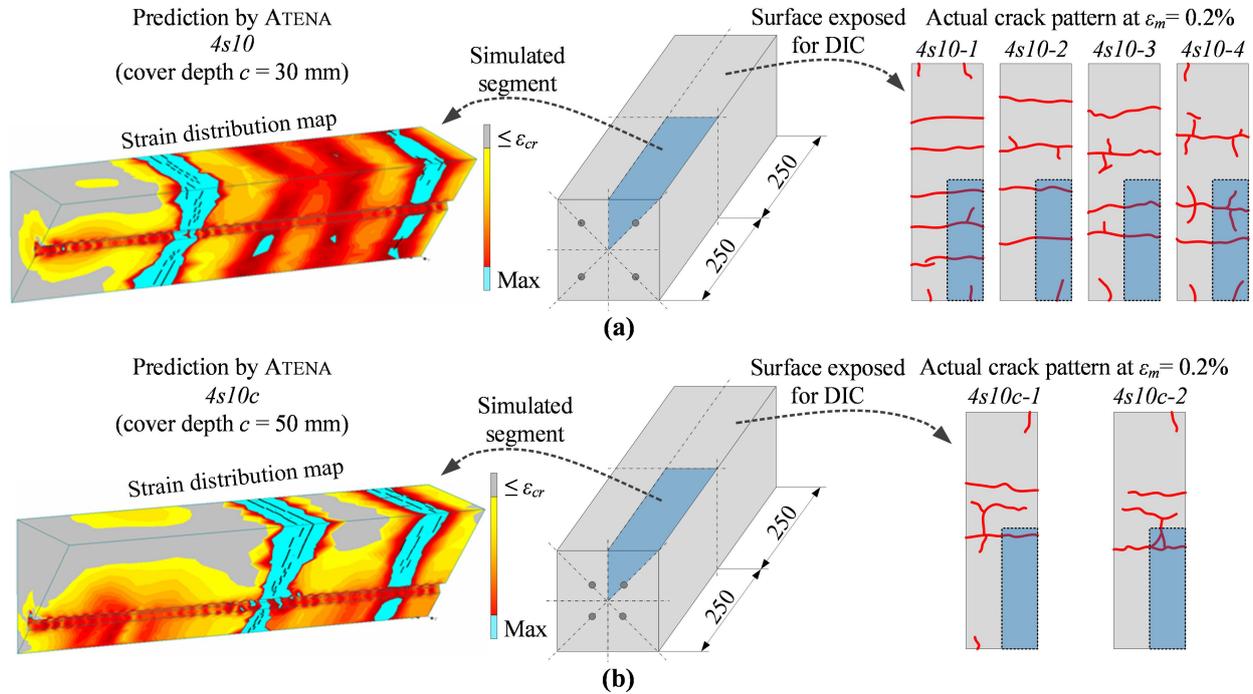

**Figure 10.** Crack patterns identified using the DIC system and predicted strain distribution in the concrete segment of the prisms with different cover depth *c*: (a) *c* = 30 mm and (b) *c* = 50 mm. (Note: the results correspond to the average deformation of the reinforcement $\varepsilon_m$ = 0.2%; in the strain distribution map, strains below the theoretical cracking value are shown in grey)

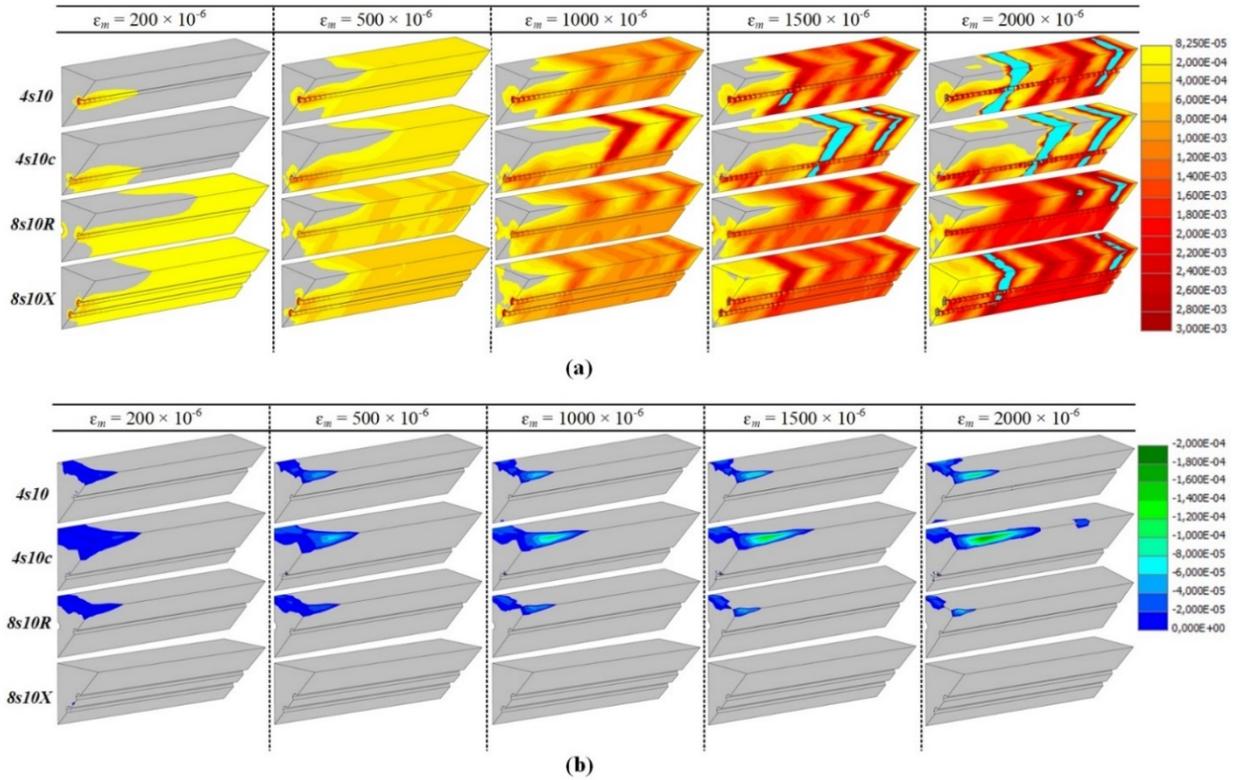

**Figure 11**. Simulated deformation behavior of the concrete segments: (a) tensile strains and (b) compressive strains

3.4 The bond modeling effect

As it was mentioned in Section 3.2, the considered bond model does not enable simulation of the radial component of the bond stresses. Such simplification was found suitable for modeling of the prism reinforced with a center bar [2,16]. It is also acceptable for modeling of the prisms *4s10* and *8s10R* with the reinforcement bars located near the perimeter of the section (Figures 9a and 9c). However, this simplification is evidently inadequate for simulating deformation behavior of the prisms *4s10c* and *8s10X* (Figures 9b and 9d) reinforced with bars spaced closely to the center of the section. To solve this problem, the ribbed bond model is included into the comparisons.

The prism *8s10X* is chosen as the object of the analysis, since the deformation predictions of this configuration of element were less adequate among all considered specimens (Figure 9). The corresponding numerical model of bond is shown in Figure 12. In this figure, the regular cylindrical bond model described in Section 3.2 is referred to as "*Model-1*", while the alternative (ribbed) model is designated as "*Model-2*". The ribbed model represents a more realistic geometry of the reinforcement bar as shown in Figure 12a. The numerical problem related with the implementation of this model could be related with FE mesh generation, since the FE size must correspond to the rib geometry. At the same time, the computation capabilities limit the number of FE. The *Model-2*, therefore, was constructed by applying several hierarchical levels of the mesh refinement. As can be observed in Figure 12b, two FE corresponds to each of the reinforcement ribs. The FE size increases with the distance from the bar. The corresponding numerical model consists of more than 225,000 FE. This model was realized at the server-workstation, since such big number of FE exceeds computation capabilities of a typical personal computer. The loading and the boundary conditions described in Section 3.2 were the same for both models. It should be noted that the *Model-2* utilizes the constant bond characteristics (designated as to "normal bond" in

Figure 8b) throughout entire length of the reinforcement bar because the ribs have been physically simulated.

Figure 13 shows results of the simulation based on both bond modeling approaches. It is evident that the application of the ribbed bond model enables almost perfect representation of average deformations of the concrete surface: minor differences are characteristic only for early crack formation stage. These inaccuracies might be addressed by representation of the inherent concrete structural defects. However, such modeling must imply stochastic information [27] that significantly complicates realization of the computation procedures. Regarding the average deformations of the reinforcement, however, as observed from the deviations of the respective curves in Figure 13 show that neither of the models yielded accurate prediction. This limitation could be attributed to the spatially highly nonlinear deformation behavior of the reinforcement bar at the extremity of the concrete prism, whereby possibly causing distortions when the average deformations are considered. Michou et al. [16] have experimentally identified this effect in concrete prism reinforced with a center bar using a distributed sensors methodology; it was also adequately simulated by using the regular bond model (the same as the *Model-1*). Modeling of behavior in the damage zone around a reinforcement bar at the extremities of a concrete block is not a trivial problem itself [37]. The present results indicate that the multiple bars configuration particularly complicates such modeling, and further research is recommended.

The adequate prediction of the surface deformations (Figure 13a) of the *Model-2* makes it an acceptable reference for analysis of the cracking problems. The crack formation character represented by the ribbed model (Figure 13b) is well agreed with the experimental crack patterns shown in Figure 6. The characteristic cracking aspect could be related with an increased length of the concrete near the extremity of the prism where surface cracks are not formed reducing the distances between the primary cracks (observed at the concrete surface). Particularly, this effect is evident in the prism *8s10X-2* (Figure 6).

The presented examples demonstrate that the well-tailored and rationalized numerical model calibrated with carefully collected test data is capable of assessing strain distribution within the concrete of a tension members reinforced with multiple bars. The simulation results can be helpful for identifying the actual stress-strain behavior of the concrete, which could not be estimated directly by using the existing experimental methods. Moreover, the numerical simulations would enable assessing the effects of bar reinforcement arrangement on the internal cracking and deformation characteristics of the concrete. Particularly, the numerical outcomes reveal limitations of the "effective area" concept in analysis of concrete members under tension. Application of such over-simplified approach to the design of complex structures in real-life might lead to noticeable errors.

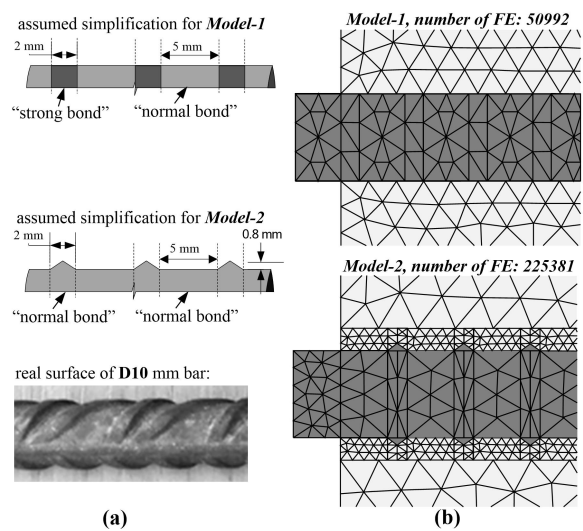
(a)  (b)

**Figure 12.** Reinforcement ribs modeling approaches: (a) a view on the models and real ribs and (b) the zoomed rib segments of the corresponding FE models

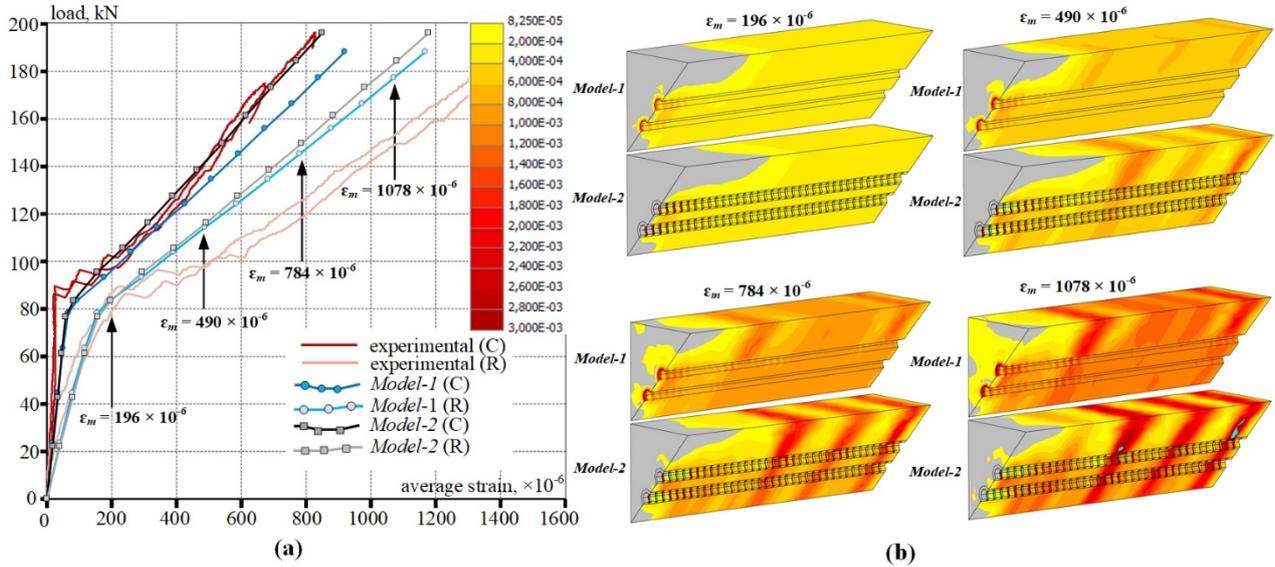

**Figure 13.** Simulation results of the specimens *8s10X* using the alternative rib models: (a) load-average strain diagram and (b) tensile strain propagation in the concrete segments

The authors would like to point out the fact that the objective of the present paper was to evaluate the strain distribution in concrete, rather the proposing a versatile bond model. In this respect, the simplified bond modelling technique proposed by Michou et al. [16] was adopted, replacing real ribbed geometry of the bars by a sequence of cylinders with the same diameter, but different surface bond properties. The further research should investigate the bond behavior of multiple bars; the effect of bar diameter on the strain localization in the concrete must be clarified as well.

## 4. Conclusions

This study has introduced new methodology for estimating strain distribution in concrete prisms reinforced with multiple bars under tension. The methodology encompasses a combination of an innovative testing setup that ensures uniform distribution of the applied tension load in multiple bars and a tailor-designed bond modeling approach for rigorous finite element analysis. The investigation has involved nominally identical 500 mm long concrete prisms reinforced with four or eight 10 mm bars distributed in a different manner in a 150 mm square cross-section. The experimental program has included 10 prismatic elements. The same specimens have also been analyzed using the finite element method based on a well-tailored regular bond model adapted from the literature. Following the symmetry conditions, the 1/16 part of the specimen was modeled in the three-dimensional domain. The proposed methodology is able rigorously analyzing the effect of arrangement of the reinforcement bars on strain distributions results. To simulate precisely the localized stress-strain states around bar ribs, the deformation behavior was also modeled by using a ribbed bond model. Such technique is requiring a huge amount of finite element and could be considered only as an academic example. The presented examples demonstrate that:
1. Deformation behavior of the concrete is dependent on the arrangement of the reinforcement. This dependency is evident from both the experimental and numerical results. Furthermore, the average deformations of the concrete are, in general, different from the average deformations of the reinforcement. Increase of the concrete cover from 30 mm to 50 mm reduces the average deformations of the concrete by 24%. This reduction is closely related with the efficiency of concrete in tension.
2. The finite element approach, based on the regular bond model and concrete softening law, calibrated with carefully collected experimental data is capable to predict deformations and cracking behavior of concrete elements reinforced with multiple bars spaced at relatively large distances. The prediction accuracy is independent on the concrete cover depth.
3. In some cases, however, neither of the considered approaches was capable to represent adequately deformation behavior of the tensile specimens reinforced with multiple (eight) closely distributed bars. This limitation could be mainly attributed to the spatially highly nonlinear deformation behavior of the reinforcement bar in the boundary zone of the concrete prism. This effect should be the object of future researches.

**Acknowledgement**

The authors gratefully acknowledge the financial support provided by the Research Council of Lithuania (Research Project S-MIP-17-62). Dr Pui-Lam Ng wishes to express his gratitude for the support provided by the European Commission under the Marie Skłodowska-Curie Actions Fellowship (Project No. 751461).

**Declaration**

The authors declare no conflict of interest with the works reported in this paper.